# Generation, Accumulation, and Utilization of Data in the Automatically Controlled End-Station of a Soft X-ray Beamline

Yasunobu Ando[a], Masashige Miyamamoto[b], Keita Hiromori[b] ‡, Takeru Nakashima[a], Masafumi Horio[b], Masato Kotsugi[c], Iwao Matsuda[b]

[a]*Institute of Integrated Research, Institute of Science Tokyo, 4259 Nagatsutacho, Midori-ku, Yokohama, 226-8501 Kanagawa, Japan*
[b]*The Institute for Solid State Physics, The University of Tokyo, 5-1-5 Kashiwanoha, Kashiwa, 277-8581 Chiba, Japan*
[c]*Department of Materials Science and Technology, Tokyo University of Science, Tokyo, 125-8585, Japan*
‡ *Corresponding author: hiromori@issp.u-tokyo.ac.jp*

## Abstract

Artificial intelligence (AI) is rapidly transforming materials exploration, yet reliable AI-assisted research depends on experimental data that retains their context, provenance, and machine-readable structure. High-throughput synchrotron measurements therefore require more than automated data acquisition: data generation, accumulation, and utilization must be connected within a continuous workflow. Here, we present a data-centric research framework that integrates the Process Informatics OperaNdo Experiment for Electronic Research (PIONEER) system, the Open Materials-data Network Entry System (OMNES), and machine-learning analysis. The PIONEER system automates sample handling, vacuum transfer, and scanning X-ray absorption spectroscopy at the BL08U soft X-ray beamline of NanoTerasu, systematically generating large volumes of spatially resolved spectral data. Measurement and analysis files are stored in the cloud-based ARIM-mdx data system, while OMNES manages metadata describing samples, preparation procedures, measurement conditions, instrument settings, data locations, and analysis histories. Tabular and JavaScript Object Notation formats are used according to the metadata structure, with numerical values stored separately from units to facilitate machine processing. OMNES organizes these records based on causal relationships among experimental and analytical processes, linking samples, measurements, raw data, processed data, and analysis results through a relational data model and visualizing their relationships as a graph. This causality-based organization preserves experimental and analytical provenance and clarifies how each result was derived from the preceding processes and data. The accumulated spectra can subsequently be processed using machine-learning methods, including dimensionality reduction and materials classification, and the resulting data can be registered back into OMNES. Previously reported machine-learning analysis of boron nitride XAS spectra is presented as an example of the analytical workflow that can be applied to data acquired at the beamline. By combining high-throughput measurement, causality-based data management, and machine-learning analysis, this framework provides a practical foundation for efficient spectral-data utilization and future AI-assisted experimental planning in materials research.

# I. INTRODUCTION

Since the onset of the third AI boom—as represented by the emergence of AlexNet in 2012 [1]—machine-learning technologies, particularly deep learning, have advanced rapidly, and techniques for processing and analyzing vast amounts of data have become widespread [2,3]. Furthermore, the 2022 launch of ChatGPT has made generative AI based on large language models easily accessible through a chat interface to not just specialists but also the general public [4]. We are currently in the midst of what might be termed the fourth AI boom, witnessing a major technological transformation comparable in scale and significance to the "Industrial Revolution"—an "AI Revolution."

The most critical factor in leveraging data-science fields such as AI and machine learning is the availability of usable data [5]. Indeed, the development of AI was historically propelled by ImageNet, a public database providing a massive collection of images [6]. However, sheer volume is not enough; to be effectively utilized in information technologies such as machine learning, the data also must be of high quality [7] and diverse [8]. In other words, AI-based information processing requires the organization and storage of vast amounts of high-quality data, a process that necessitates the coordinated execution of three stages: data generation, data accumulation, and data utilization [5].

While conducting cutting-edge research in condensed matter science [9–13], we have implemented data-driven approaches within the research environment, focusing on fundamental physics experiments involving the optical absorption of materials. Using X-rays as the probe beam, it is possible to identify the constituent elements of a material based on the photon energies where the absorption coefficient changes dramatically, which are known as absorption edges [14,15]. Furthermore, the spectrum near these X-ray absorption edges—referred to as the X-ray absorption fine structure—reflects the material's electronic state; because the spectral shape is highly dependent on the material, it can serve as a "fingerprint" for material identification [14,15]. We developed a system to automate these X-ray absorption spectroscopy (XAS) experiments, enabling the unstaffed acquisition of large volumes of high-quality data. By integrating this with automated analysis using machine learning, we have pioneered a new approach to material characterization.

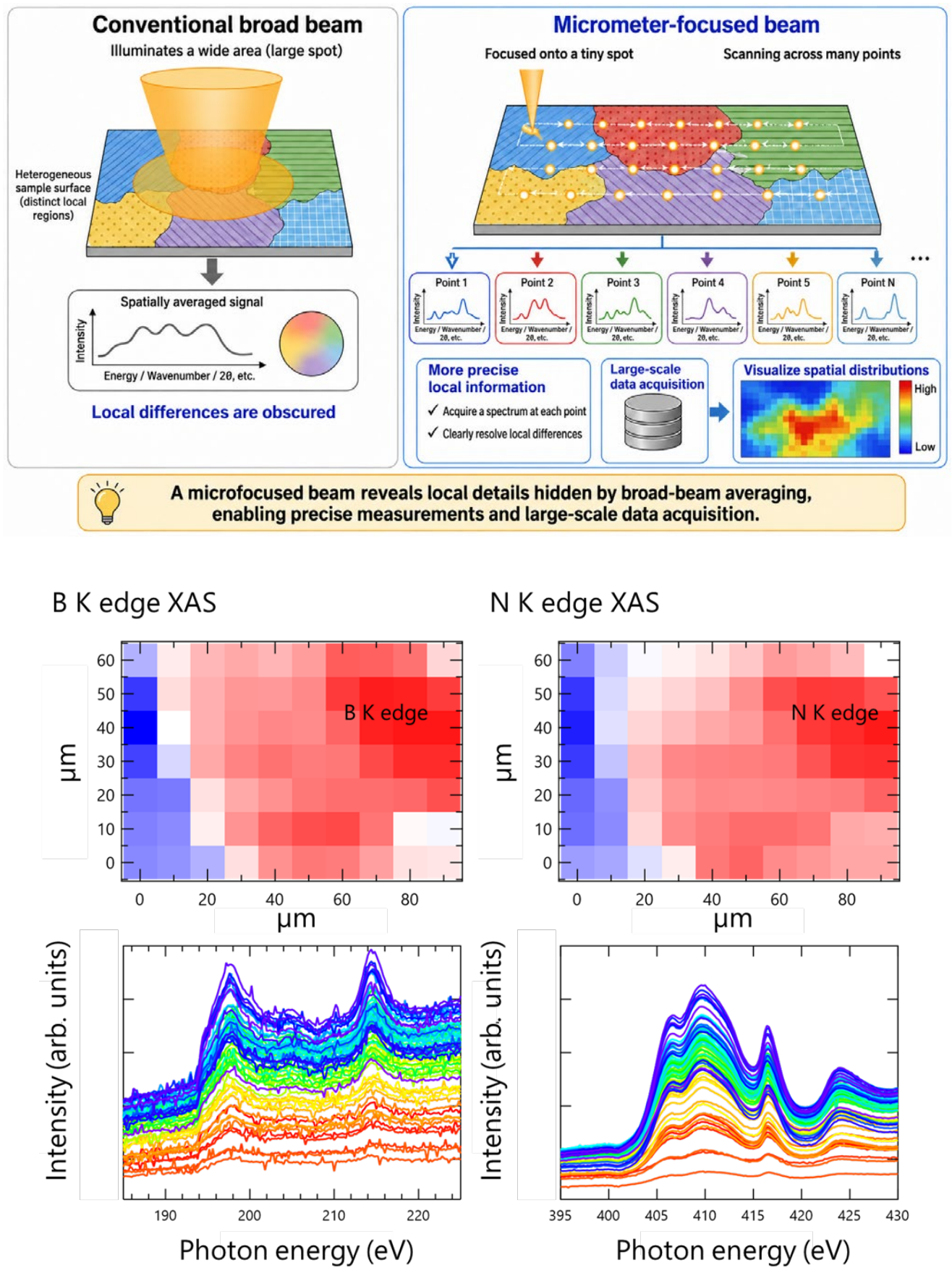

Figure 1: (Top) Method for data collection by scanning measurements using a micro-beam. (Bottom) Sets of experimental XAS spectra of cubic boron nitride (c-BN).

# II. Data Generation: Fully Automated Experimental System with Robots

This system was developed for the BL08U soft X-ray beamline at NanoTerasu, a synchrotron radiation facility located in Sendai City, Miyagi Prefecture, Japan [16]. Because this beamline utilizes a high-brightness X-ray beam, sufficient absorption signals can be detected within a short period. Furthermore, using X-ray mirrors to focus the beam into a tiny spot, it is possible to perform localized XAS while scanning the beam's position [17–19]. We currently generate micro-beams on the order of micrometers tailored to specific objectives, enabling pinpoint measurements of numerous tiny samples scattered across a surface and the spectral mapping of heterogeneous sample surfaces [17]. Whereas measurements using conventional, large-spot X-ray beams (Figure 1, top left) yield only averaged sample data, the use of micro-spots (Figure 1, top right) collects localized information reflecting the sample's actual characteristics, thereby enabling more accurate material evaluation. This XAS not only reveals the distribution of constituent elements and electronic states in principle but also enables the acquisition of information regarding local crystal structures through fingerprint analysis. As a representative example, Figure 1 bottom panel also shows spatially resolved B *K* edge and N *K* edge XAS spectra of cubic boron nitride (c-BN) acquired at different measurement positions.

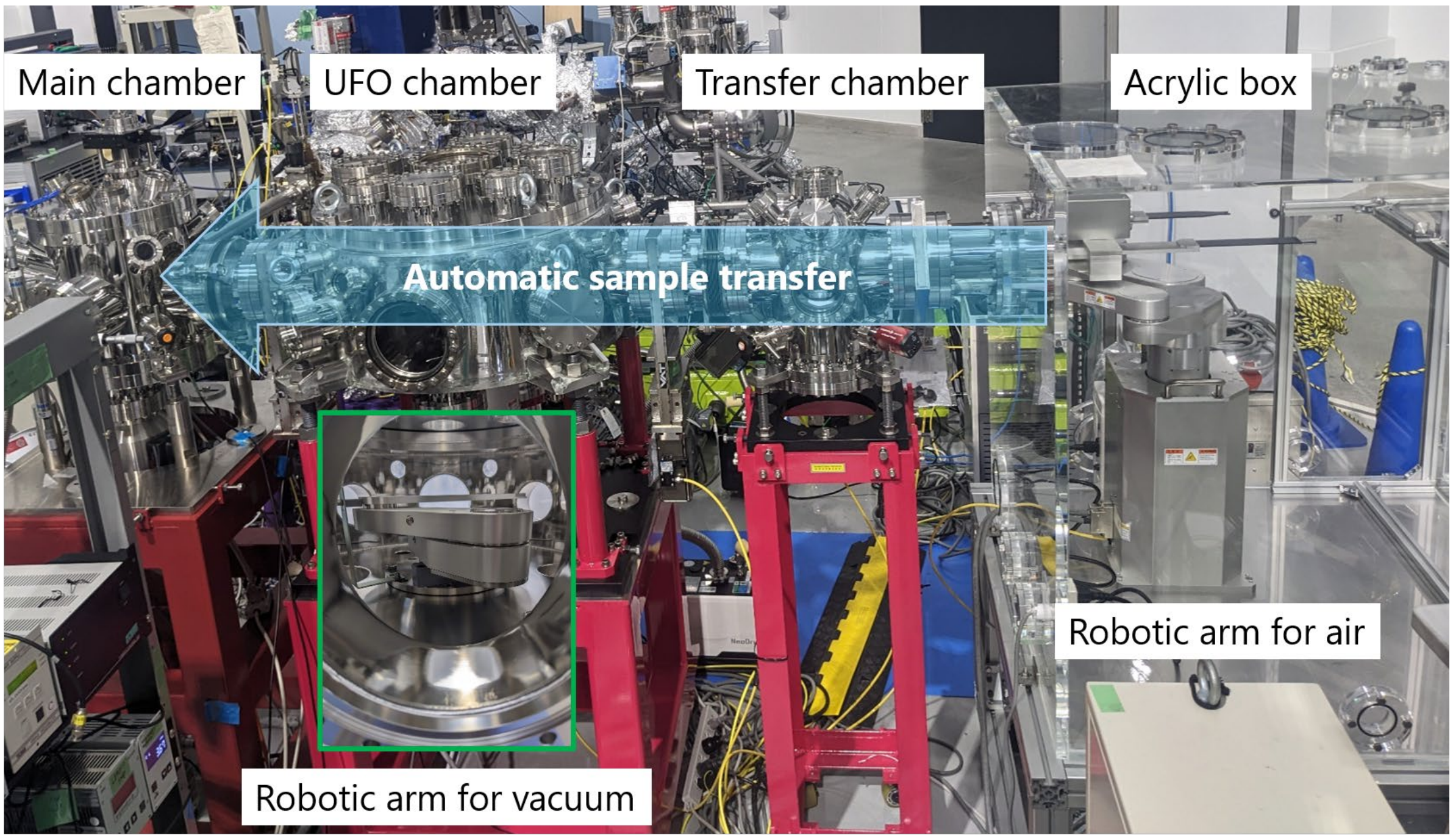

Figure 2: Process Informatics OperaNdo Experiment for Electronic Research (PIONEER) system, which we developed for the BL08U beamline at the NanoTerasu synchrotron radiation facility.

To conduct such X-ray absorption experiments, we integrated a robot into the apparatus to automate the entire data collection process, from sample transfer to measurement. Figure 2 shows the developed system, which is called the Process Informatics OperaNdo Experiment for Electronic Research (PIONEER) system. From right to left, an acrylic box, a stainless steel (SUS) transfer chamber, an SUS UFO (vacuum) chamber, and an SUS main (vacuum) chamber are shown. The probe light used in our XAS method consists of soft X-rays, which have relatively long wavelengths, and is subject to significant absorption by atmospheric molecules [14,15]. Consequently, the main chamber is maintained at a level of vacuum comparable to that of outer space ($10^{-7}$ Pa). A robotic arm installed inside the acrylic box handles both sample preparation and the transfer of samples between the transfer and vacuum chambers. The transfer chamber features an automatic internal pressure adjustment system; once a sample is placed inside, a vacuum pump activates to evacuate the chamber to a vacuum level of $10^{-5}$ Pa. Subsequently, a vacuum-compatible robotic arm located in the UFO chamber grasps the sample and transfers it from the transfer chamber to the main chamber (Figure 2 inset, bottom). Gate valves separate the sample preparation, transfer, preparation, and main chambers, opening and closing based on the vacuum status of each chamber and movements of the robotic arms. Computer systems manage and control the robotic arm movements, gate valve operations, and internal pressure; this enables the coordinated transfer of samples from atmospheric to vacuum conditions and the automatic execution of scanning XAS measurements. Currently, we are able to acquire tens of thousands of data points (approximately 100 spectra) daily using fully automatic operations simply by placing various samples into the UFO chamber.

## III. Data Accumulation: Metadata and Database

To preserve the vast amount of data generated by the PIONEER system, it is crucial to consider its organization and storage, as well as which specific data should be preserved. Simply saving the measurement data without organization would hinder the ability to locate and utilize them. Furthermore, since measurement and analysis results can vary depending on factors such as the type and composition of the sample, sample preparation conditions, and instrument settings, it is necessary to store information regarding these experimental conditions alongside the measurement data themselves. The key to this process is the use of *metadata*, which are data used to describe the data [20,21].

In measurements conducted by the PIONEER system, the metadata include information such as the sample name, material composition, sample preparation conditions, date and time of measurement, synchrotron radiation energy, measurement location, equipment settings, data storage location, and analysis methods. Without this information, it is impossible to distinguish whether differences in the data stem from the material itself or from measurement conditions and equipment settings. Similarly, when training an AI, failure to accurately link measurement data with information such as material composition and chemical state increases the risk that the AI learns incorrect features. Therefore, it is essential to manage and store measurement data and metadata together as an integrated unit [22]. Furthermore, data and metadata storage require accessibility and availability. To achieve this, it is preferable to organize the data into a *database*, which is a structured collection of data. The simplest form of data structuring is a single-file format. Tabular formats, which classify data into rows and columns, are the most widely used for storing metadata; the Comma-Separated Values (CSV) format is a common example. As the name implies, a CSV file represents data columns by separating the data elements in each row using the "," symbol (a comma) [23]. In contrast, JavaScript Object Notation (JSON) files are well-suited for storing hierarchical or tree-structured data. JSON files allow nested data structures, making it possible to represent tree-like data arrangements [24].

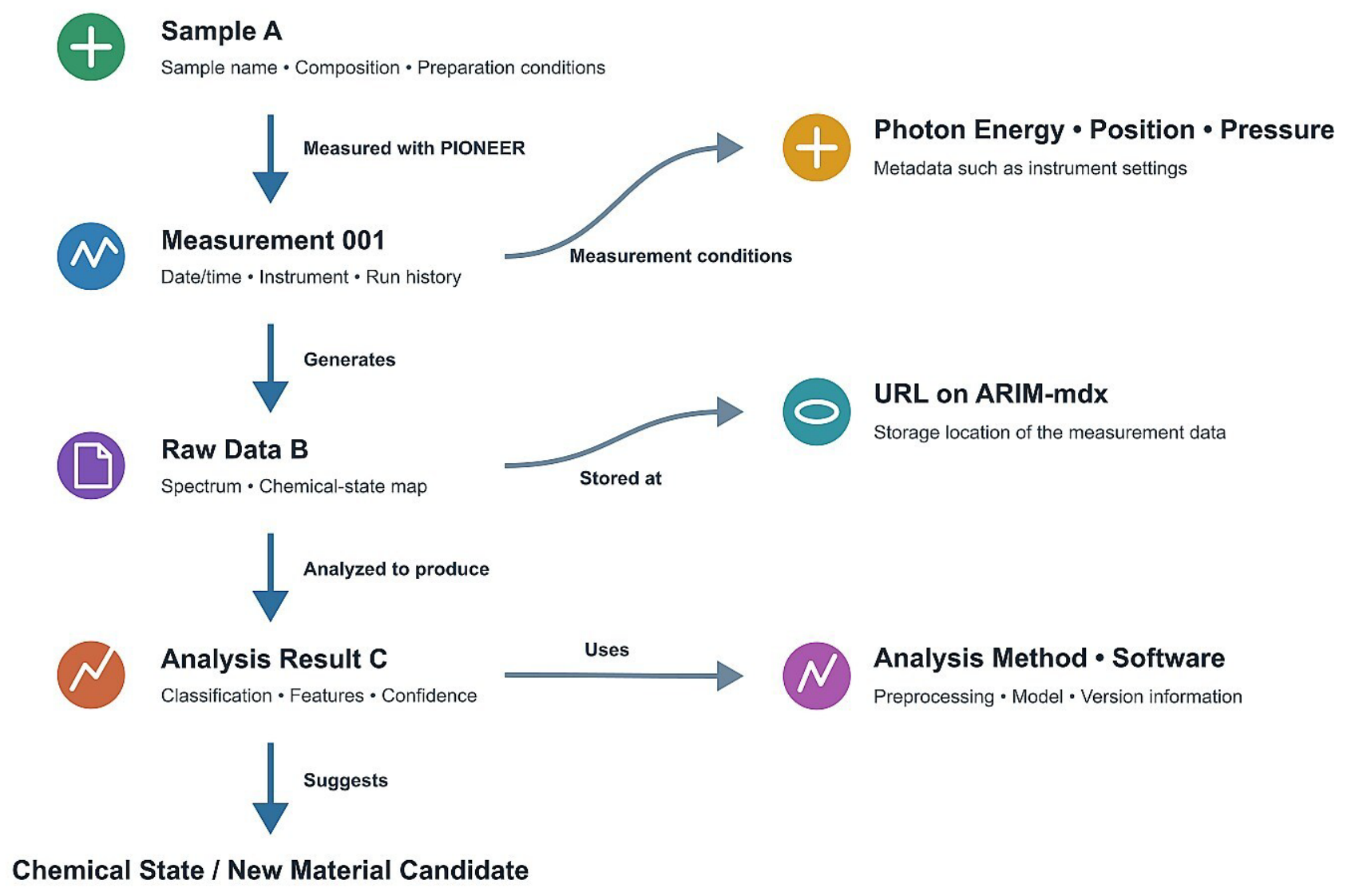

Figure 3: Management of measurement data and metadata by OMNES. For the actual data stored in ARIM-mdx, the relationships between them are recorded in OMNES.

When storing numerical metadata in CSV, the numerical value and its unit should be placed in separate fields[25]. Combining a number and a unit in a single data element makes automated reading and processing more difficult. JSON can represent the same separation within nested objects. In the following example, metadata describing the measurement conditions and the sample are grouped independently:

```
{
  "MeasurementMetadata": {
```

```
    "DataURL": "https://...",
    "Slit": {
      "unit": "mm",
      "value": 20
    }
  },
  "SampleMetadata": {
    "SynthesizedDate": "2026-04-01",
    "Ingredients": {
      "Material A": {
        "unit": "mL",
        "value": 100
      }
    }
  }
}
```

As in CSV, storing values separately from units and field definitions helps prevent errors when the data are reused.

The primary measurement and analysis data are stored in the cloud-based ARIM-mdx data system operated by the University of Tokyo [26]. In addition, a separate metadata database has been constructed to organize these data. We use the Open Materials-data Network Entry System (OMNES) database to organize and store metadata (such as sample details, measurement conditions, raw data, analysis methods, analysis results, and storage locations) within ARIM-mdx in either tabular or JSON format. By focusing on metadata in this way, a key advantage is the ability to provide a unified structure even if the underlying raw data formats are not standardized. Furthermore, the relationships between the stored metadata are organized and maintained across multiple tables. A database structured with such links is known as a relational database [27].

OMNES is also excellent at visualizing data relationships. As research progresses, one sample is measured many times under different conditions, and one set of measurement data is analyzed using multiple methods, increasing the complexity of the relationships between the data, which grow like the branches of a tree. Therefore, OMNES uses a graph database to express and visualize the connections between data based on the information accumulated in a relational database. In a graph database, elements such as samples, measurements, and analysis results are represented as *nodes* (points), and their relationships are represented as *edges* (lines) [28–30]. This allows a user to intuitively check the flow of an experiment using the connections between points and lines; for example, measure sample A with PIONEER → generate raw data B → examine using analysis method C → find a specific chemical state (Figure 3).

This graph structure is designed primarily based on the causal relationships between data. Nodes represent files or the data themselves, while edges can be interpreted as measurement methods or the operation of measuring. In other words, an edge acts as a "function," and a graph structure expressed as node A → node B can be understood to mean that the data stored in node A is "transformed" into the data stored in node B by the operation contained in the edge. For example, in a measurement experiment, a target sample exists, and this can be interpreted as the cause that generates the measurement data. Similarly, for computational data, an output file is obtained from an initial structure file through the operation of calculation. Most research data inherently encapsulates this causal structure.

Thus, OMNES enables the entire experimental process to be recorded through metadata. In recent years, there have been discussions regarding the public release of databases used in research, leading to the proposal of the Findable, Accessible, Interoperable, and Reusable (FAIR) data principles [31,32]. OMNES, which transforms experimental records into a database, can serve as a foundation for data storage that satisfies the FAIR principles. Furthermore, the system allows specific analysis results or generated figures to be traced back to the original measurement data, conditions, and samples; this facilitates the investigation of experimental procedures or conditions that may have influenced new discoveries. Looking to the future, integrating OMNES with generative AI or AI agents could enable AI to search existing data, compare them, and propose which samples should be measured next and under what locations or conditions [33,34]. Our vision for data-centric materials research involves "generating" data with PIONEER, "accumulating" it in ARIM-mdx and OMNES, and "utilizing" it through AI and other technologies.

## IV. Utilizing Data: Machine-Learning Analysis

Implementing robotics in the developed system has yielded several benefits: 1) robots can operate in vacuum environments where humans cannot; 2) precise robotic movements ensure high task reproducibility; 3) by conducting experimental procedures previously performed by humans, robots allow researchers to focus on more creative activities [35,36]; and 4) high-quality data can be stably and systematically generated and accumulated for use in AI-driven information processing databases. Because the PIONEER system generates vast amounts of data, manual processing is time-consuming, and manual data analysis can be subject to individual variability.

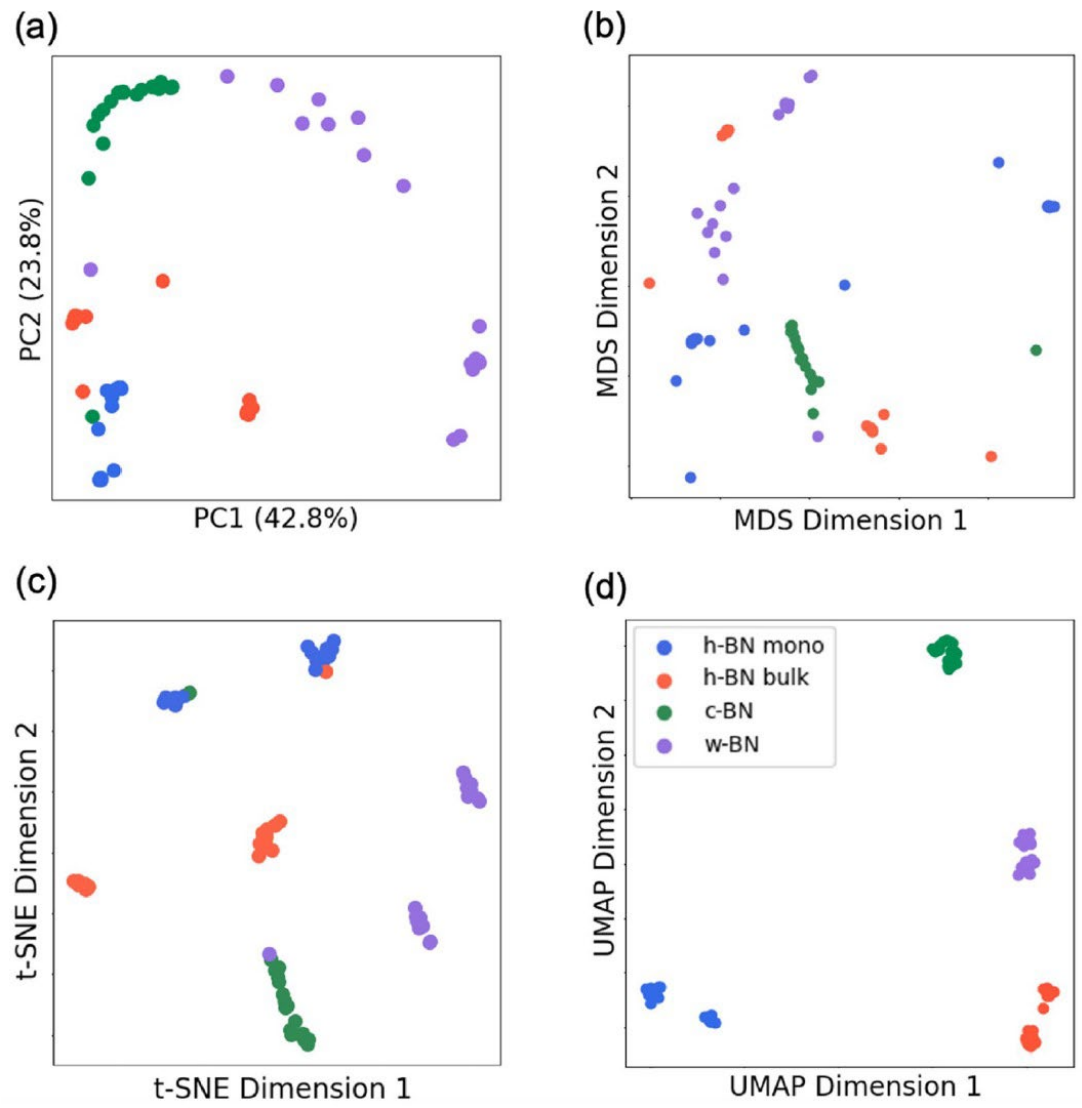

Figure 4: Machine learning analysis (dimensionality reduction methods) of the X-ray absorption spectra of boron nitride (BN). (a) Principal component analysis. (b) Multidimensional scaling (MDS). (c) t-distributed stochastic neighbor embedding (t-SNE). (d) Uniform manifold approximation and projection (UMAP). Reproduced from [37] under the CC BY 4.0 license.

To address these issues, we employ automated analysis using machine learning; Figure 4 illustrates an example of this approach. The figure shows the results of analyzing calculated X-ray absorption spectra of boron nitride (BN) using various dimensionality reduction methods [37]. In this paper, these dimensionality reduction methods are treated as "black boxes," and only the analysis results are compared. While the resulting spectral groups are classified within their respective parameter spaces, the uniform manifold approximation and projection machine-learning analysis (Figure 4(d)) clearly distinguishes between BN crystal structures (h-BN, c-BN, and w-BN). Focusing on h-BN, the analysis further classifies samples into monolayers and bulk crystals, demonstrating that information regarding film thickness has also been obtained. When the PIONEER system was used to perform actual XAS on the h-BN, c-BN, and w-BN samples, the clusters for the experimental data agreed well with those of the calculated data for the corresponding crystal structures (Figure 5). In other words, this automated machine-learning analysis can be used for material identification. Hence, synthesized or unknown samples can be evaluated with high throughput, thereby accelerating the pace of research. By contrast, if experimental data points fall outside the classification of existing substances, it signifies the discovery of a new substance; this holds great significance because it has the potential to lead to the formulation of new scientific principles and the development of innovative materials in the field of materials science.

The data generated through these machine-learning analyses are also stored in the OMNES cloud database, where they are organized and managed in a relational database format with interlinked records. Subsequently, these data will be further utilized for more advanced analyses and applications. In future, the PIONEER system, which integrates these three processes of data generation, accumulation, and utilization, will be linked with generative AI and AI agents to enable even more advanced capabilities.

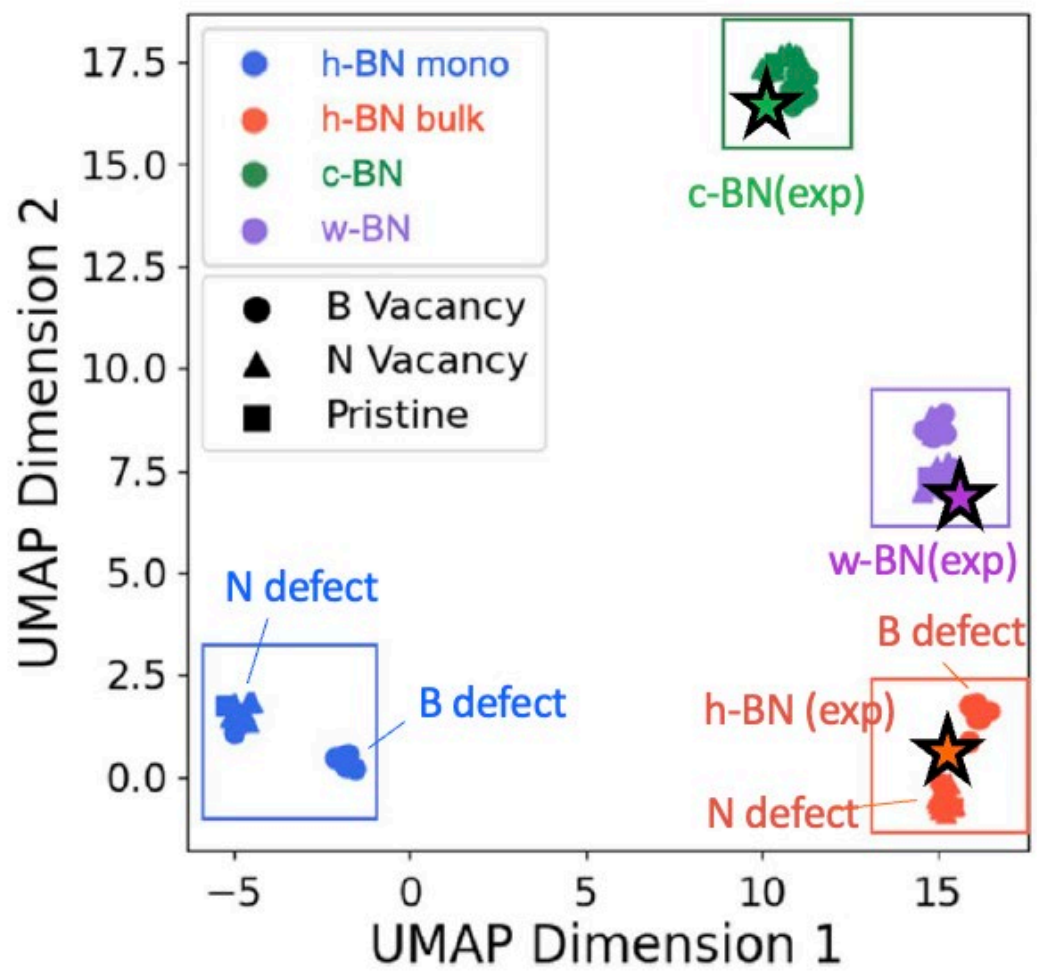

Figure 5 Example collection of experimental and calculated X-ray absorption spectra of various boron nitride (BN) crystals after UMAP machine-learning analysis. Data measured by the PIONEER system are indicted by stars. Calculated data are from [37] courtesy of M. Kotsugi.

## V. Conclusion

Reliable, high-quality data are essential for the advancement of physics—a field dedicated to exploring the fundamental truths of nature—making the collection, management, and analysis of such data critically important. The seamless integration of AI and robots into research environments where data are generated, stored, and utilized not only significantly boosts research efficiency but also enables AI to provide researchers with new insights through methods such as the machine-learning analysis of large amounts of data. In future, the collaboration between humans and AI/robotics is expected to drive new research developments, leading to increasingly unique and diverse experiments in the field of physics.

### Acknowledgments

This work was supported by JST CREST (Grant JPMJCR21O4), Japan. This work was also financially supported by the New Energy and Industrial Technology Development Organization (NEDO) of Japan through a project for the development of common platforms to promote the widespread utilization of hydrogen (JPNP25002). We thank Kimberly Moravec, PhD, from Edanz (https://jp.edanz.com/ac) for editing a draft of this manuscript.

## Graphical abstract

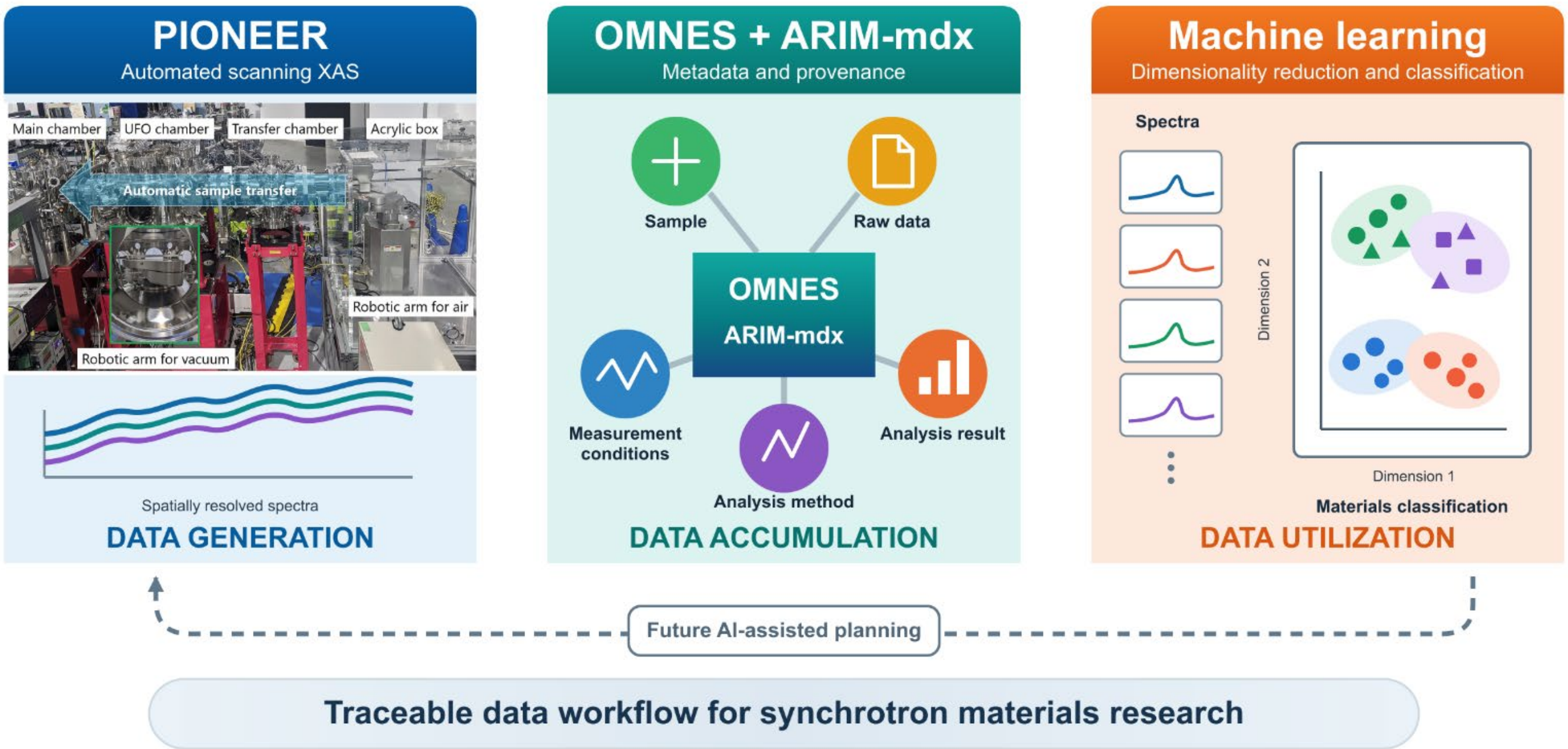